\documentclass[a4paper]{jpconf}
\usepackage{graphicx}
\usepackage{amstext}

\begin{document}
\title{The XENON10 WIMP Search Experiment at the Gran Sasso Underground Laboratory}

\author{Laura Baudis\footnote{On behalf of the XENON collaboration}}

\address{RWTH Aachen University, Physics Department, Sommerfeldstr. 14, 52074 Aachen, Germany}

\ead{laura.baudis@rwth-aachen.de}

\begin{abstract}

XENON10 is a new direct dark matter detection experiment using liquid xenon as target for weakly interacting, 
massive particles (WIMPs). A two-phase (liquid/gas) time projection chamber with 15\,kg fiducial mass has been installed 
in a low-background shield at the Gran Sasso Underground Laboratory in July 2006. After initial performance tests with 
various calibration sources, the science data run started on August 24, 2006. The detector has been running stably since then, and a
full analysis of more than 75 live days of WIMP search data is now in progress. We present first results on gamma and neutron  
calibration runs, as well as a preliminary analysis of a subset of the WIMP search data. 

\end{abstract}

\section{Introduction}

The goal of XENON \cite{xenon} is to search for interactions of massive, cold dark matter particles in liquid xenon.
The motivation of this search comes from our current understanding of the universe.
Over the last ten years, a variety of cosmological observations, from the primordial abundance of light
elements, to the study of large scale structure, to the observations of high redshift supernovae, to the detailed 
mapping of anisotropy of the cosmic microwave background, have led to the construction of a so-called 
concordance model of cosmology. In this  model, the universe is made of $\sim$4\% baryons which constitute the ordinary
matter, $\sim$23\% nonbaryonic dark matter and $\sim$73\% dark energy \cite{spergel}.  Understanding the nature of dark matter poses 
a significant challenge to astrophysics. The solution 
may involve new particles with masses and cross sections characteristic of the electroweak scale.
Such Weakly Interactive Massive Particles (WIMPs), which would have been in thermal equilibrium with quarks and leptons in the hot 
early universe and decoupled when they were non-relativistic, represent a generic class of dark matter candidates~\cite{lee1977,jungman}. 
If WIMPs are the dark matter, their density in the Milky Way halo may allow them to be detected in laboratory experiments by looking for the nuclear 
recoils produced in elastic WIMP--nuclei collisions \cite{goodman}. A WIMP with a typical mass between 10\,GeV --10\,TeV will 
deposit a nuclear recoil energy below 100\,keV  in a terrestrial detector. The expected rates are determined by the WIMP-nucleus cross section and 
by their  density and velocity distribution in the vicinity of the solar system \cite{jungman}.
Direct-detection experiments, in particular CDMS~II~\cite{cdms_prl06,cdms_prd06}, {\small CRESST}~\cite{cresst}, {\small EDELWEISS}~\cite{edelweiss},  
{\small ZEPLIN~II}~\cite{zeplin} and {\small WARP}~\cite{warp}, are beginning to significantly constrain the WIMP-nucleon scattering cross section 
and, for the first time, start to probe the parameter space which is predicted by supersymmetric extensions to the Standard Model (recent reviews can be found 
in \cite{gabriel,rick,laura}).

While it is clear that cryogenic experiments such as CDMS are still leading the field, detectors based on liquid noble elements (Ar, Xe) are rapidly evolving 
and are already catching up in sensitivity. 
Liquid argon (LAr) and xenon (LXe) have excellent properties as dark matter targets. They are intrinsic scintillators, with 
high scintillation ($\lambda$ = 128~nm for Ar, $\lambda$ = 175~nm for Xe) and ionization yields. They are available in large quantities 
(with LAr being much cheaper) and can be purified to 1\,ppt (parts per trillion)-levels. 
Scintillation in LAr and LXe is produced by the formation of excimer states, which are bound states of ion-atom systems. 
If a high electric field ($\sim$1 kV/cm) is applied, ionization electrons can also be detected, either directly or through the secondary process of 
proportional scintillation. Measuring both the primary scintillation signal and a secondary process yields a method of discriminating between electron 
and nuclear recoils. In LAr, an additional differentiation, namely the time difference between the decay of the singlet and triplet excited states 
(6 ns versus 1.6\,$\mu$s) is being used.
An advantage of LXe is its high density (3\,g/cm$^3$),  which provides self-shielding and allows  for compact detectors, and  
the high atomic number (Z=54, A=131.3), which is favorable for scalar WIMP-nucleus interactions.
Moreover, since natural Xe has two isotopes with spin ($^{129}$Xe, $^{131}$Xe, at the combined level of almost 50\% abundance), 
a liquid xenon detector is susceptible for both coherent and axial-vector WIMP-nucleus couplings.

\section{The XENON10 Experiment}

XENON10 is a new direct dark matter search experiment with the aim of observing the small
energy released after a WIMP scatters off a xenon nucleus \cite{xenon}. 
The  simultaneous detection of ionization and scintillation in a liquid xenon (LXe) 3D position
sensitive time projection chamber (TPC) allows to distinguish nuclear recoils (as produced by WIMPs and 
neutrons)  from the dominant electron recoil background (as originated from photon and electron interactions) \cite{elena1,elena2}.
The detector has been transported from the Nevis 
Laboratory at Columbia University to Gran Sasso in March 2006 and  has been installed 
in its low-background shield in mid July 2006. It has been taking data continuously and very stably since then, 
with a total of more than 75 live days of WIMP search data accumulated.

\begin{figure}[ht!]
\begin{center}
\includegraphics[width=18pc]{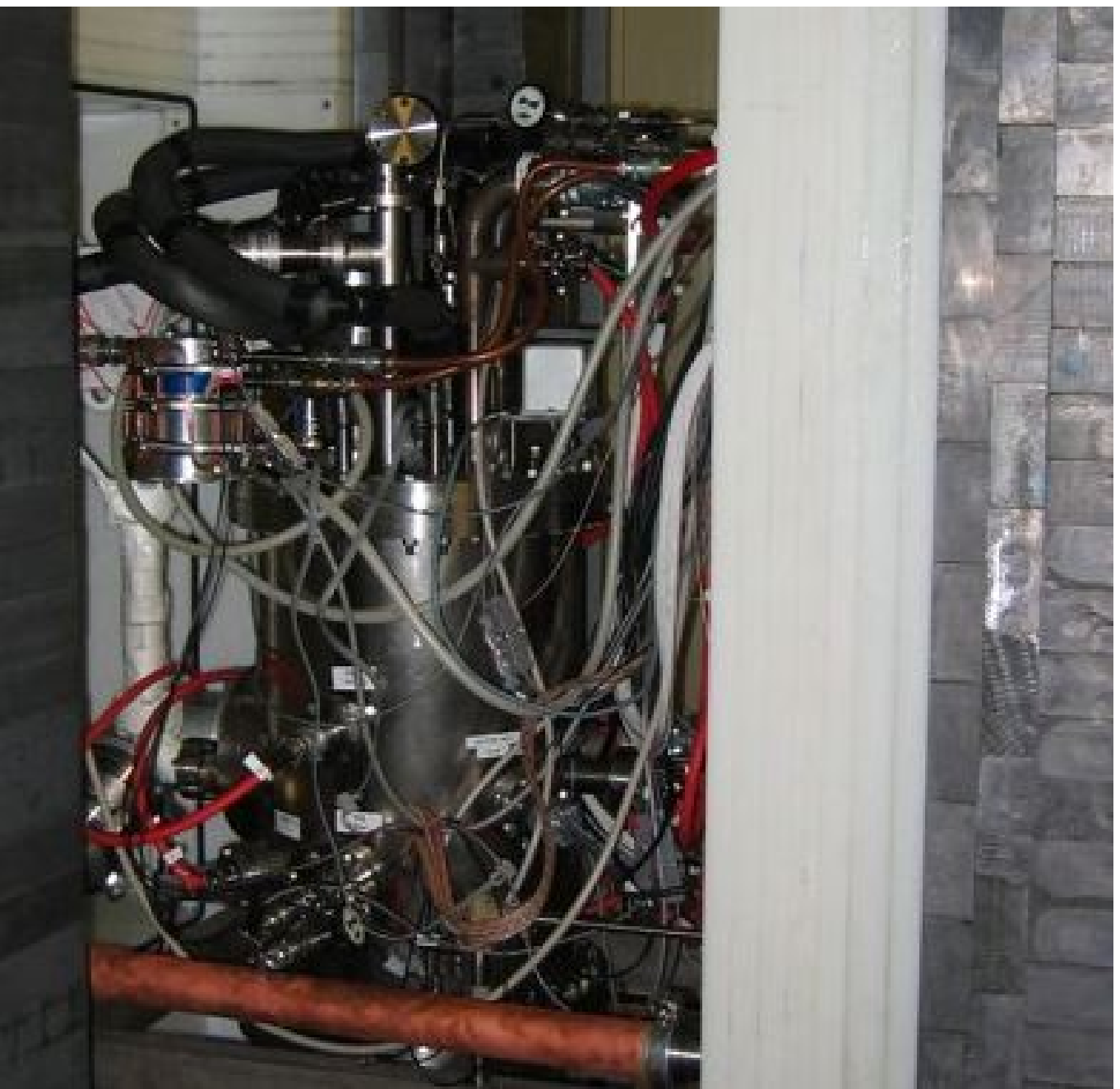}
\includegraphics[width=18pc]{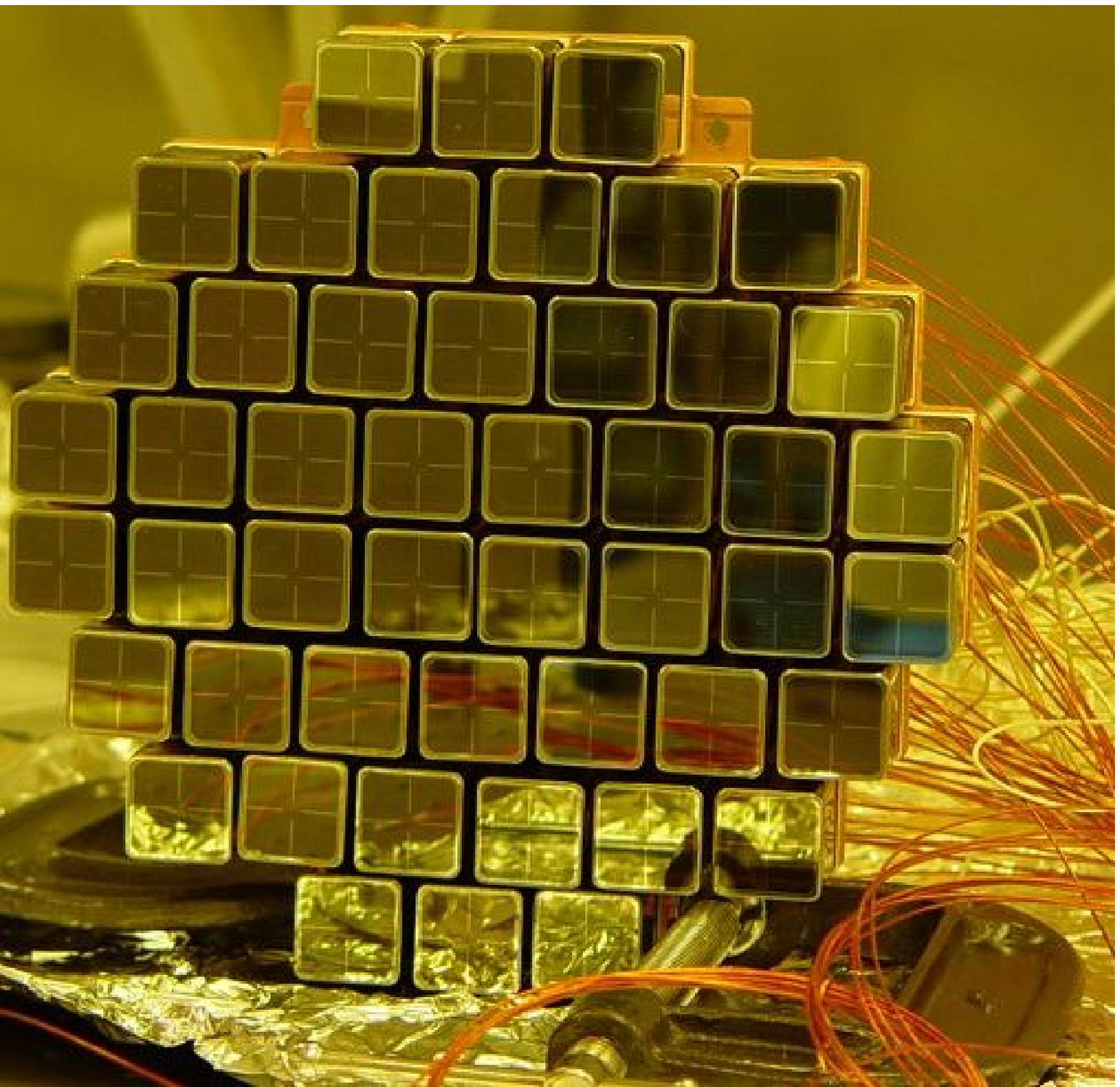}
\end{center}
\caption{\small (left) A view of the XENON10 TPC in its low-background shield during the initial installation. 1.6 tons of 
polyethylene and 34 tons of lead are used to reduce the external neutron and gamma background to insignificant levels. 
(right) A picture of the top photo-multiplier array (48 Hamamatsu R8520-AL PMTs) before its insertion into the inner detector chamber.}
\label{fig:xenon10}
\end{figure}

\begin{figure}[ht!]
\begin{center}
\includegraphics[width=18pc]{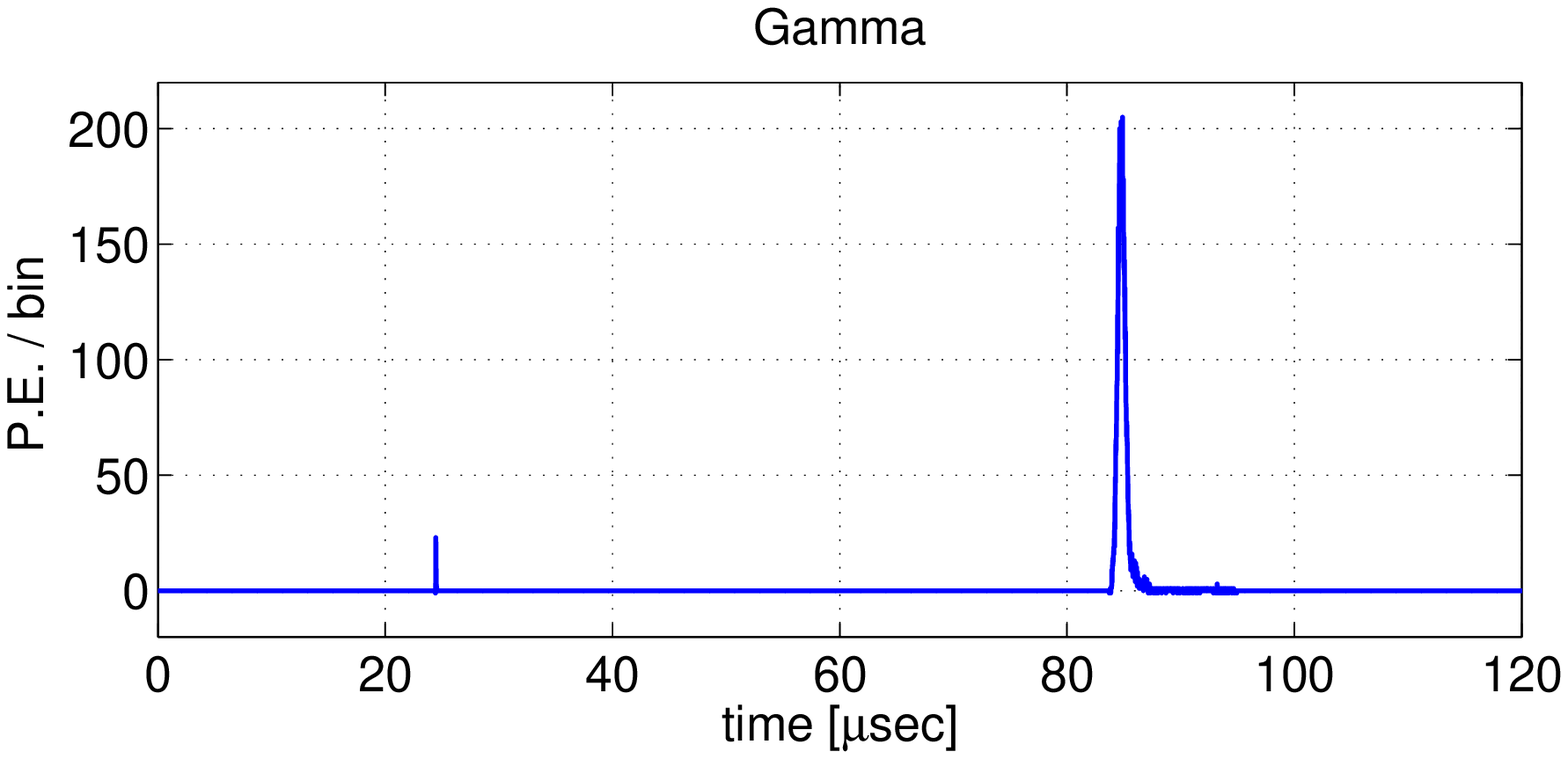}
\includegraphics[width=18pc]{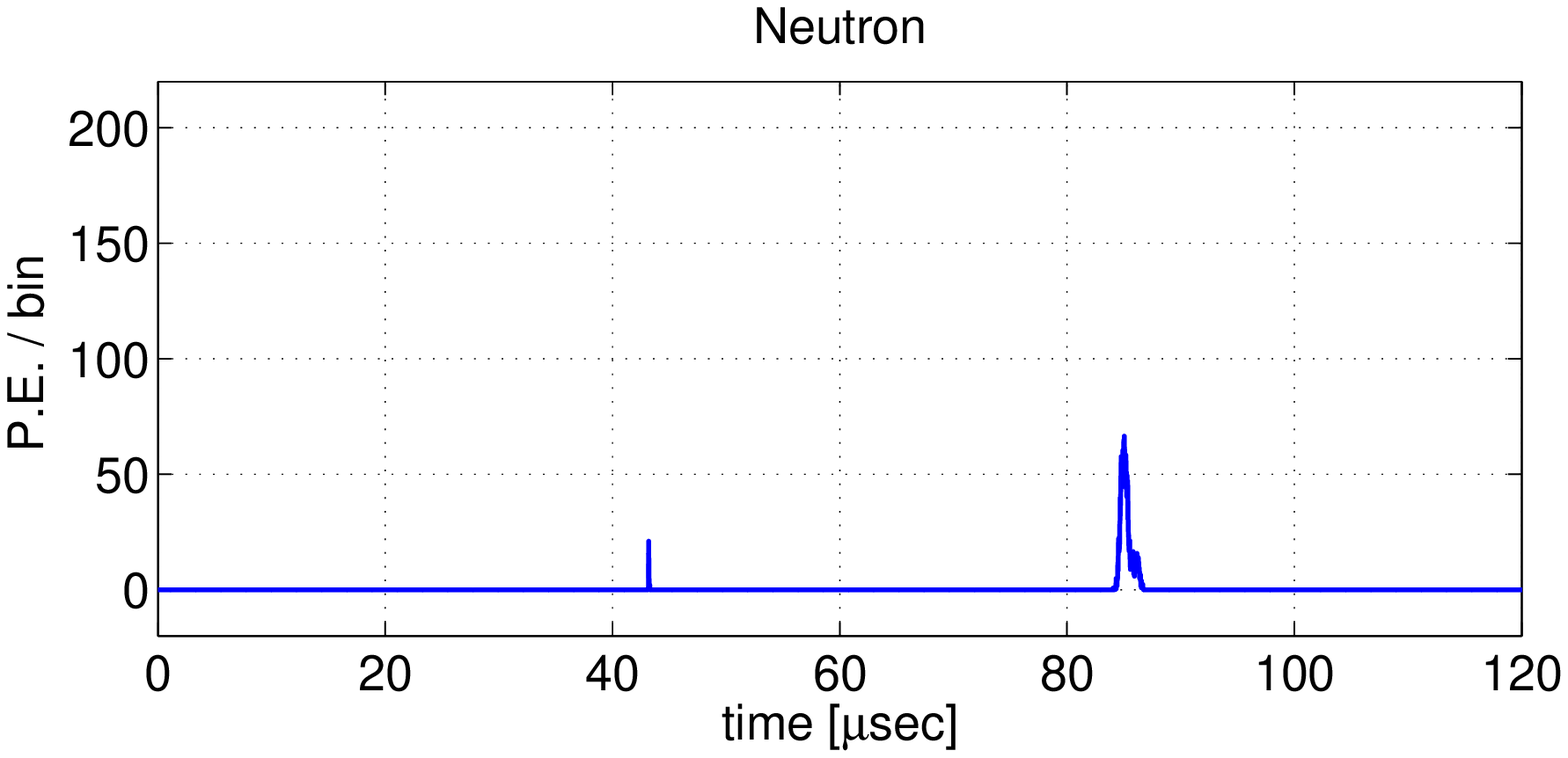}
\end{center}
\caption{\small  (left) Low-energy electron recoil event (photon induced), with a S1 signal of 70 photoelectrons and a S2 of 
1.7$\times10^4$ p.e.
(right) Low-energy Xe nuclear recoil event (neutron induced), with a S1 signal of 70 photoelectrons and a S2 of 5.3$\times10^3$ p.e.}
\label{fig:events}
\end{figure}

The dual-phase (liquid and gas) TPC is filled with 22\,kg of ultra-pure liquid xenon, the active 
volume being defined by a teflon cylinder with an inner diameter of 20\,cm and a 
height of 15\,cm (the total active mass is 15\,kg of LXe). 
The TPC is equipped with four wire meshes, two in the liquid and two in the gas phase. 
The bottom mesh serves as a cathode and the next one, positioned just below the liquid level, forms, 
together with a series of field shaping rings, the 15\,cm  drift region. The two last meshes, along with the one 
below the liquid level, serve to define the gas proportional scintillation region.
The active xenon volume is viewed by 89 Hamamatsu R8520-06AL 1\,inch square PMTs, 35\,mm high. 
The bottom array of 41 PMTs is located below the cathode, fully immersed in LXe, and mainly detects the 
prompt light signal (in the following labeled S1). The 48 PMTs of the top array are located in the cold gas above the liquid, detecting the proportional 
light signal (labeled S2) which is created by the collision of extracted electrons with Xe atoms in the gas phase (the drift field is 0.7\,kV/cm).
Figure \ref{fig:xenon10} shows the XENON10 detector in its low-background shield (left) as well as a picture of the top PMT array (right).
The hit pattern in this array is used to reconstruct the x-y position of an event with few mm resolution. The z-coordinate is 
calculated from the time difference between the pulses of direct and proportional light (with a  maximum of 75\,$\mu$s and a resulting 
z-resolution of $<$1\,mm).  Typical low energy electron and nuclear recoil events detected by XENON10 PMTs are 
shown in Figure \ref{fig:events}. As we will demonstrate in the next sections, the 3D position sensitivity and the self shielding of LXe,  serves, 
along with the prompt versus proportional light ratio, as a fundamental background rejection feature. 

The cooling of XENON10 is provided by a  pulse tube refrigerator directly in contact with the Xe gas.  
We have attained a very stable pressure ($\Delta P < 0.006$ atm) and temperature ($\Delta T < 0.005 ^\circ$C)
for over six months of operation. Under these conditions the fluctuation of PMT gains is $<2\%$. 
To reach the high purity required for a LXe TPC with a drift gap of 15\,cm, we are continuously 
circulating the xenon gas through a high temperature getter.  The electron
lifetime inferred from recent data is $(1.8 \pm 0.4)$\,ms which corresponds to a $\ll 1 \, \text{ppb}$
(O$_2$ equivalent) xenon purity.

\section{XENON10 {\it in situ} Calibration Measurements} 

\begin{figure}[ht!]
\begin{center}
\includegraphics[width=18pc]{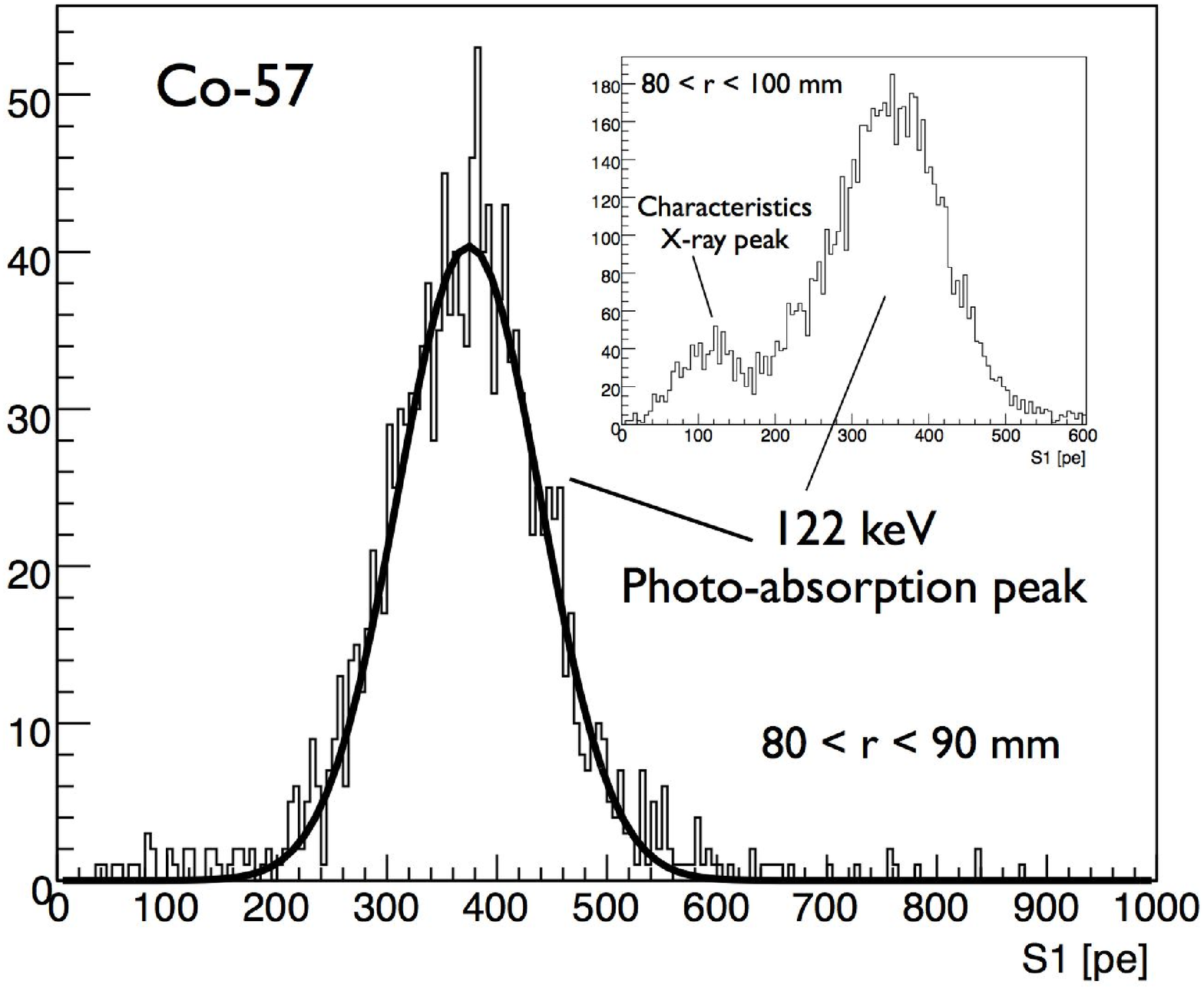}
\includegraphics[width=18.5pc]{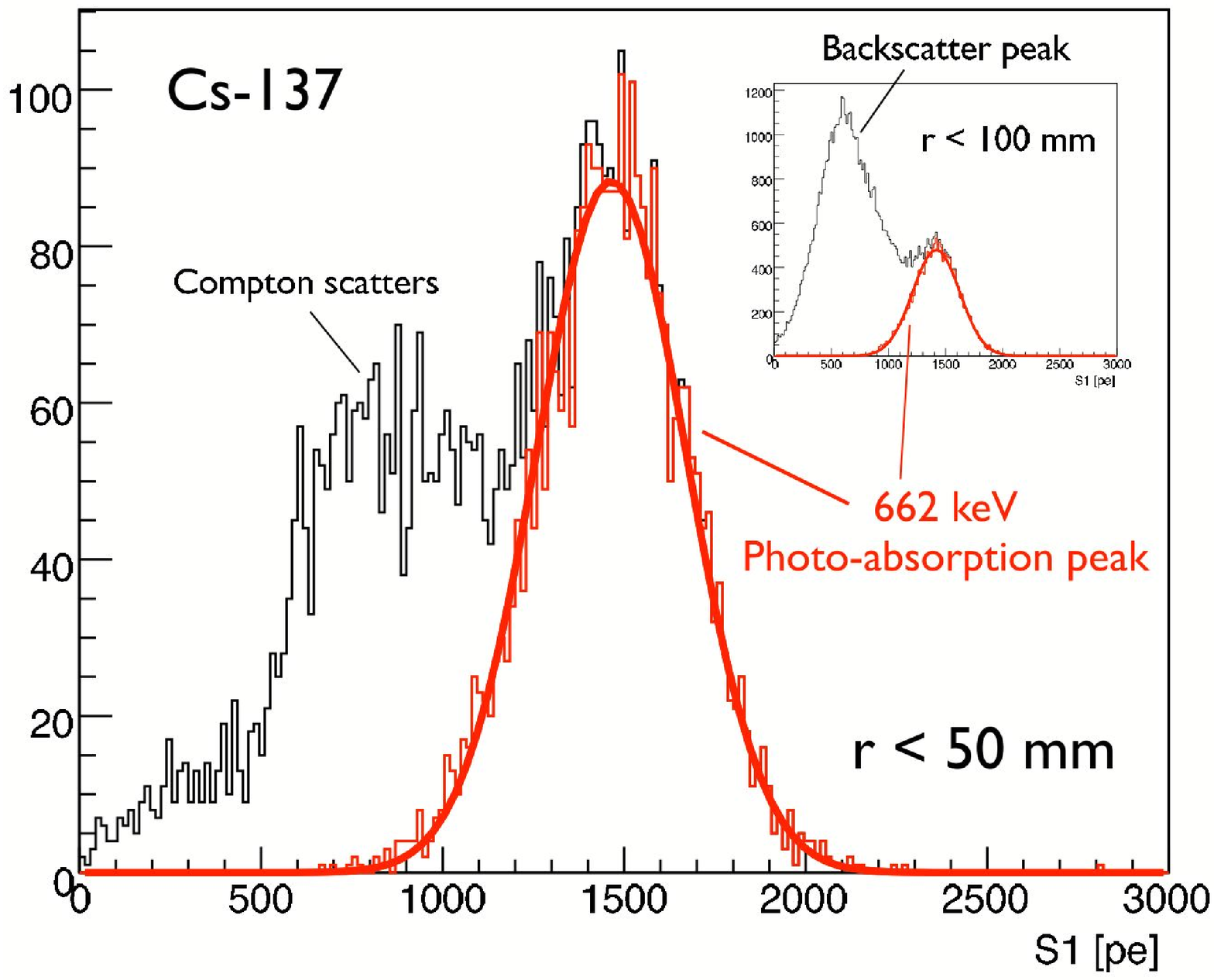}
\end{center}
\caption{\small (left) S1 scintillation spectrum from a $^{57}$Co calibration. The light yield for the 
122\,keV photo-absorption peak is 3.1 p.e./keV. (right) 
S1 scintillation spectrum from a $^{137}$Cs calibration. The light yield for the 662\,keV photo-absorption peak is 2.2 p.e./keV.}
\label{fig:calib}
\end{figure}

We have used gamma ($^{57}$Co, $^{137}$Cs)  and neutron (AmBe) calibration sources, as well as neutron-activated xenon 
($^{131m}$Xe, $^{129m}$Xe) to extensively calibrate the XENON10 detector at the underground laboratory.
To achieve a minimum impact on WIMP search exposure time, the passive lead and polyethylene shields were designed to allow insertion of external, 
encapsulated sources without exposing the detector cavity to outside, radon-contaminated air. 
Figure~\ref{fig:calib} (left) shows a scintillation light spectrum (S1) from a $^{57}$Co calibration. For a radius $8<r<9$\,cm 
the average light yield for 122\,keV gamma rays is 374 photo-electrons (p.e.) (3.1 p.e./keV), with an energy 
resolution  $\sigma$ of 17\%. The inset shows the clear detection of the characteristic X-ray peak from the $K$-shell at $\sim$30\,keV. 
A scintillation light spectrum from a $^{137}$Cs calibration is shown on the right side of the same figure.
The 662\,keV photo-absorption peak  yields an average light yield of 1464 p.e. (2.2 p.e./keV) and an energy resolution  
$\sigma$ of 205 p.e. (14\%). The light yield does not change substantially (less than 5\%) for radii $r<9$\,cm.

\begin{figure}[ht!]
\begin{center}
\includegraphics[width=18pc]{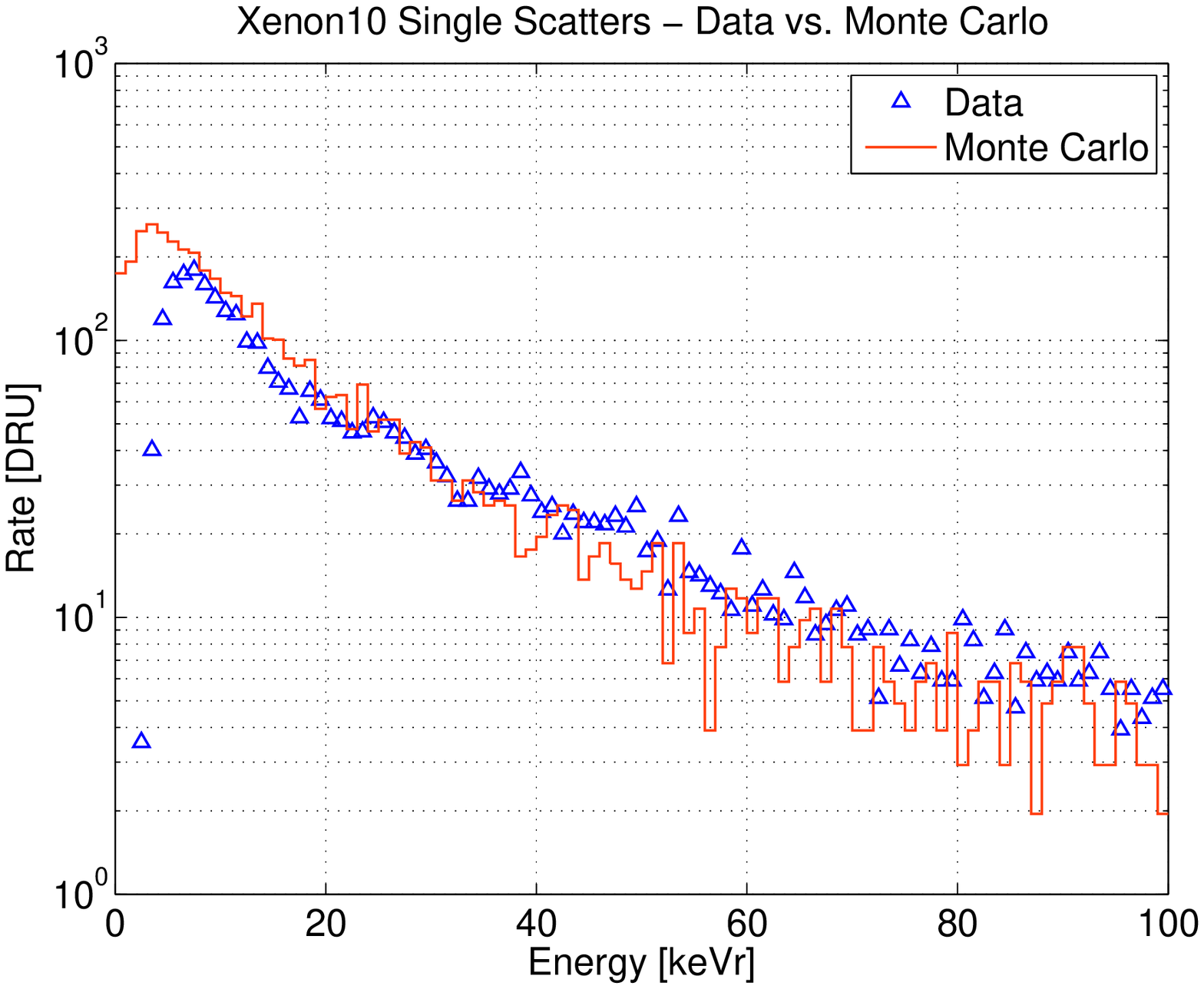}
\includegraphics[width=18pc]{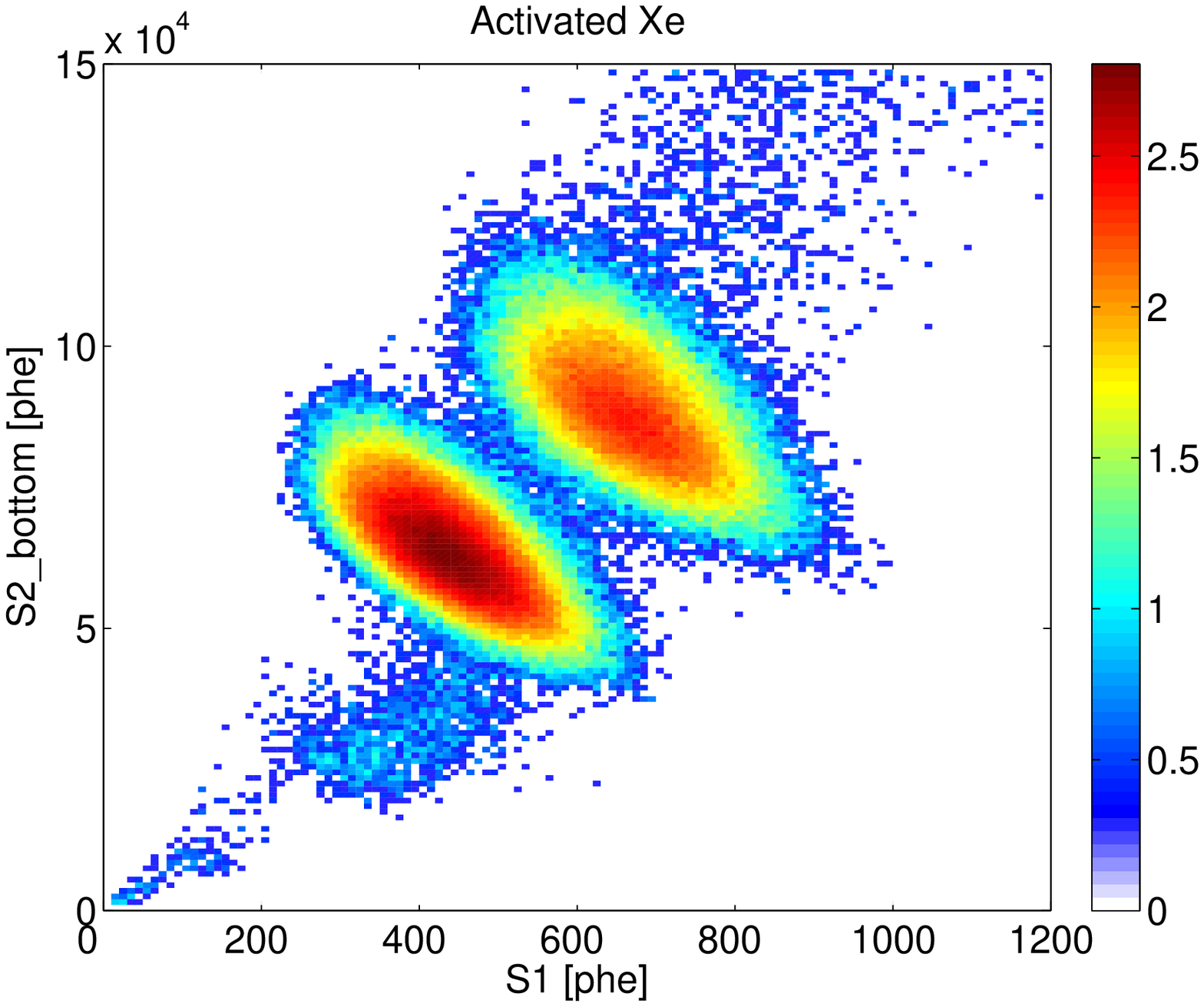}
\end{center}
\caption{\small (left) 
A data versus MC comparison of the energy spectrum of single elastic nuclear recoils in XENON10. The data was taken during an 
AmBe neutron calibration  on December 1, 2006. (right)  S2 versus S1 signal for n-activated Xe data. The gamma lines at 164\,keV and 236\,keV 
and the S2-S1 anti-correlation are clearly visible.}
\label{fig:ncalib}
\end{figure}

On December 1, 2006, we have calibrated the XENON10 detector with an encapsulated AmBe neutron source with an activity 
of 200\,neutrons/s. Such a calibration allows to determine the predicted WIMP signal region as well as the 
detector's ability to distinguish between nuclear and electron recoils down to its energy threshold. 
The energy threshold was determined to be at $\sim$10\,keV nuclear recoil energy (keVr) with a light yield of 0.7 photo-electrons/keVr. 
In Figure \ref{fig:ncalib} (left) we show a data versus Monte Carlo comparison of the energy spectrum of single elastic nuclear recoils in LXe.
We note the good agreement between the data and the Monte Carlo above $\sim$10\,keVr. 
At the time of this writing a small amount of neutron activated xenon gas was added to the XENON10 detector.   
Two xenon meta-stable states, $^{131m}$Xe and $^{129m}$Xe 
decay emitting 164\,keV and 236\,keV gamma rays with half-lives of 11.8 days and 8.9 days, respectively.
These photons allow for a more uniform energy and position calibration across the LXe active volume.
Figure \ref{fig:ncalib} (right) shows a preliminary S2 versus S1 plot for 
a subset of the activated Xe data: the two gamma lines and the anti-correlation between the S1 and the 
S2 signals are clearly visible.

\section{Preliminary XENON10 Backgrounds, Results and Near Future Plans}

\begin{figure}
\begin{center}
\includegraphics[width=18pc]{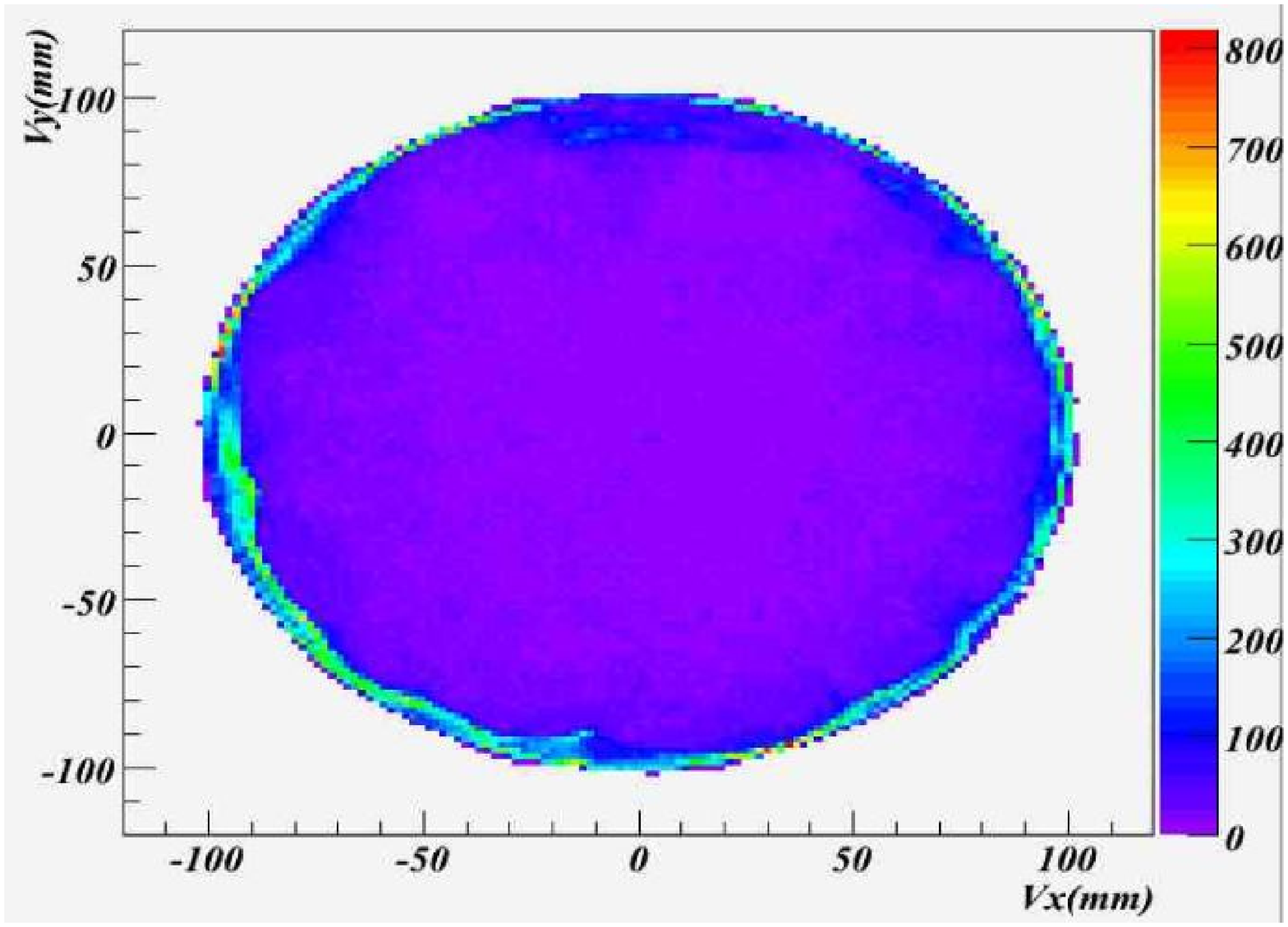}
\includegraphics[width=19pc]{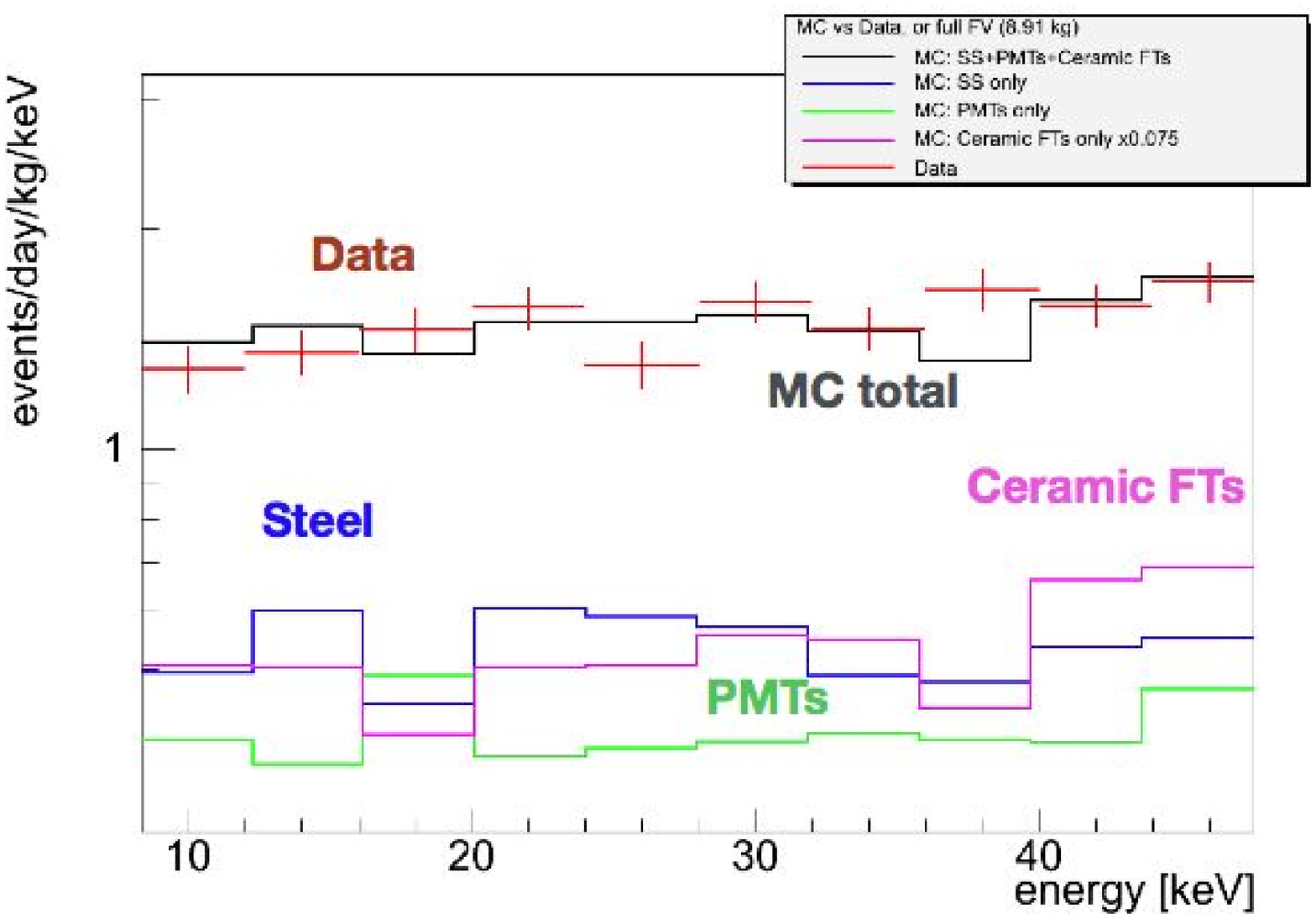}
\end{center}
\caption{\small (left) x-y position of background events for a subset of the WIMP search data. The 
shielding power of the dense liquid, and the advantage of 3-D position sensitivity are evident. 
(right) Comparison of XENON10 background data (red crosses) with Monte Carlo predictions 
(MC sum in solid black) from radioactivity in the PMTs, ceramic and stainless steel detector components.}
\label{fig:bgs}
\end{figure}

The XENON10 backgrounds are dominated by gamma interactions originating from the remaining radioactivity of detector materials.
In Figure \ref{fig:bgs} (left) we show the x-y-distribution of background events from a subset of WIMP search data. 
A radial (along with a z-axial) cut dramatically reduced the background to the level of about 1 event/kg/day/keV.  
The right side of the same figure shows a histogram of the remaining background after fiducial volume cuts, 
together with the result of a  Monte Carlo simulation based on the measured activities of the detector and shield materials. 
The backgrounds can be explained by the U/Th/K/Co radioactivity in the PMTs, 
in the stainless steel used in the detector and cryostat, and in the signal and high voltage ceramic feedthroughs.

\begin{figure}[ht!]
\begin{center}
\includegraphics[width=18pc]{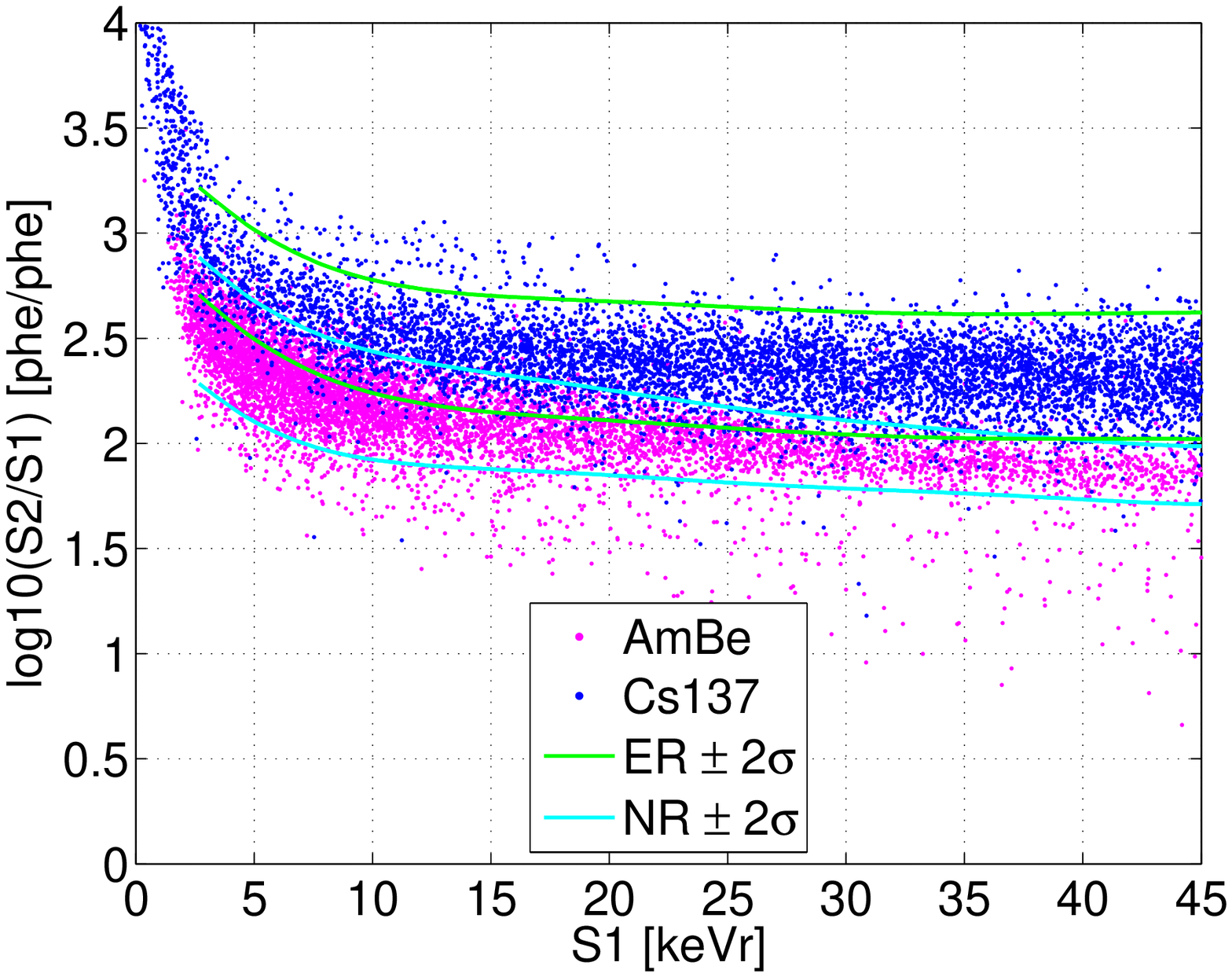}
\includegraphics[width=19pc]{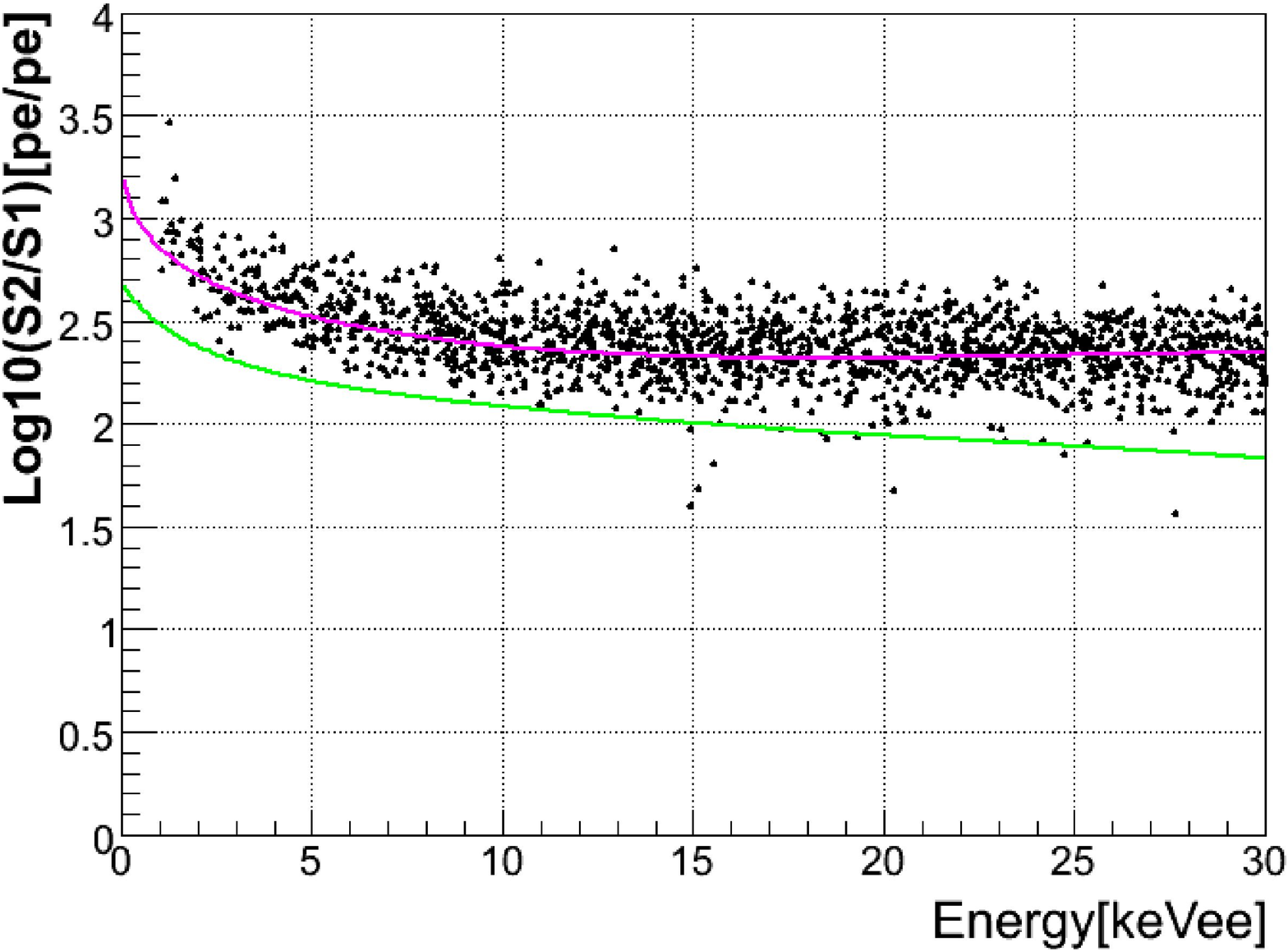}
\end{center} 
\caption{\small (left) Electron and nuclear recoil bands in the S2/S1 versus S1 parameter space (from $^{137}$Cs and AmBe data),  
and the $\pm$2$\sigma$-regions of the calculated band centroids. We remind the reader that the data is taken at a drift field of 0.7 kV/cm. (right)  
XENON10 ionization yield versus scintillation energy, for a subset ($\sim$22 live days) of the WIMP 
search data. The magenta and green lines correspond
	to the centroid of the electron and nuclear bands, as determined from calibration data.}
\label{fig:discrimination}
\end{figure}

The sensitivity to WIMP-nucleon interactions depends on the detector's ability to distinguish between electron and 
nuclear recoils, for the dominant background stems from electron recoil interactions. We have studied the XENON10 
discrimination based on  $^{137}$Cs gamma and AmBe neutron calibration data. Figure \ref{fig:discrimination} (left) shows the 
electron and nuclear recoil bands in the log$_{10}$(S2/S1) versus S1 parameter space, along with the calculated band centroids and 
the $\pm$2$\sigma$-regions. These regions were determined by fitting  Gaussian distributions to the data, divided in 5\,keV 
energy intervals.  For a nuclear recoil acceptance at the level of 50\% between 10--40\,keVr, the achieved discrimination is 99.5\%. 
We note that XENON10 is operated at a drift field of 0.7\,kV/cm, and that we expect a higher level of discrimination for a larger 
drift field, as will be realized in the upgraded XENON10 version.

The total statistics of the dark matter search data  amounts to about 75 days of
live time with a 15\,kg sensitive volume. Figure~\ref{fig:discrimination} (right) shows a plot of log$_{10}$(S2/S1) as a
function of energy for $\sim$22 live days, giving a total exposure of 38\,kg\,d after cuts. The magenta
and green lines correspond to the centroid of the electron and nuclear recoil bands, respectively. 
In the energy region 3-15\,keVe (10-40\,keVr), zero events are observed in the 50\% acceptance
window of the nuclear recoil band. In the region above 15\,keVe energy, we observed few events in
the nuclear recoil band. Our current understanding is that these are double scatter  
events, with one interaction in a passive region and one in the fiducial volume. Since no S2 signal is observed
for scatters in passive xenon, the additional S1 signal from such a region results in a
smaller log$_{10}$(S2/S1) ratio. We are currently testing this hypothesis with Monte Carlo simulations and comparison 
with calibration data, 
and are planning to block the passive xenon regions for the S1 light signal for an upgraded XENON10 version.

We are in the process of analyzing the complete science data set, whereby we have blinded about two-thirds of the  
total WIMP search data.  We are developing the data quality and physics cuts, 
and are testing their efficiencies on calibration and non-blinded WIMP search data. 
We expect that the existing configuration can probe WIMP-nucleon cross sections down 
to 1$\times$10$^{-7}$pb for coherent WIMP-nuclei interactions. 
Concomitantly, we are preparing an upgrade of the XENON10 detector.
Modest changes, such as the replacement of  ceramic feedthroughs with lower activity ones, already in hand, 
the replacement of the stainless steel vessel with a vessel made of OFHC Copper, selection of low radioactivity PMTs, etc, 
would reduce the expected background to 0.1 events/keV/kg/day, one order of magnitude
below the current level. We also expect that operation of the detector at a higher drift field  will 
improve the nuclear versus electron recoil discrimination, and that optically masking the dead xenon regions will 
decrease the number of potential leakage events.
If lower backgrounds and a higher discrimination can be achieved, XENON10 could reach a WIMP-nucleon sensitivity
of 2$\times$10$^{-8}$pb by 2008.

\section*{Acknowledgments}
This work was funded by NSF Grants No. PHY-03-02646 and No. PHY-04-00596, the CAREER Grant No. PHY-0542066, the DOE Grant No. DE-FG02-91ER40688, 
by the Volkswagen Foundation (Germany) and by the FCT Grant No POCI/FIS/60534/2004 (Portugal). We would like to thank the LNGS/INFN staff and engineers 
for their continuous help and support.

\end{document}